\RequirePackage{fix-cm}
\documentclass[natbib]{svjour3}      
\smartqed 
\usepackage{graphicx}
\usepackage{epsf}
\usepackage{amssymb}
\usepackage{amsmath}
\usepackage{placeins}
\usepackage{color}
\journalname{SSRv}
\newcommand{\be}{\begin{equation}}
\newcommand{\ee}{\end{equation}}
\newcommand{\beq}{\begin{eqnarray}}
\newcommand{\eeq}{\end{eqnarray}}
\newcommand\subsun[1]{{$_{\normalsize\odot}$}}

\newcommand{\fig}[1]{Fig.~\ref{fig:#1}}

\newcommand{\rj}{\,r_{\rm j}}

\newcommand{\sigmain}{\sigma_{\rm in}}
\newcommand{\rhot}{r_{\rm L,hot}}
\newcommand{\gammamax}{\gamma_{\rm max}}
\newcommand{\newtext}[1]{#1}

\begin{document}
\title{Magnetoluminescence}
\titlerunning{Magnetoluminescence}      
\author{R.~Blandford, Y.~Yuan, M.~Hoshino, L.~Sironi}
\institute{R. Blandford \at KIPAC, Stanford University, Stanford, CA 94305, USA \email{rdb3@stanford.edu}
\and
Y. Yuan \at Lyman Spitzer, Jr. Postdoctoral Fellow,
Dept. Astrophysical Sciences, Princeton University, Princeton, NJ 08540, USA \email{yajiey@astro.princeton.edu}
\and
M. Hoshino \at University of Tokyo, 7-3-1 Hongo, Bunkyo, Tokyo 113-0033 \\ \email{hoshino@eps.s.u-tokyo.ac.jp}
\and
L. Sironi \at Department of Astronomy, Columbia University, New York, NY, 10027 \\ \email{lsironi@astro.columbia.edu}
 }
\date{Received: date / Accepted: date}
\maketitle
\begin{abstract}
Pulsar Wind Nebulae, Blazars, Gamma Ray Bursts and Magnetars all contain regions where the electromagnetic energy density greatly exceeds the plasma energy density. These sources exhibit dramatic flaring activity where the electromagnetic energy distributed over large volumes, appears to be converted efficiently into high energy particles and $\gamma$-rays. We call this general process magnetoluminescence. Global requirements on the underlying, extreme particle acceleration processes are described and the likely importance of relativistic beaming in enhancing the observed radiation from a flare is emphasized. Recent research on fluid descriptions of unstable electromagnetic configurations are summarized and progress on the associated kinetic simulations that are needed to account for the acceleration and radiation is discussed.  Future observational, simulation and experimental opportunities are briefly summarized.
\end{abstract}
\section{Introduction and Context}
The discovery of cosmic rays over a century ago \citep{Hess:1912aa}, of double radio sources and radio supernova remnants in the 1940s \citep{Reber:1947aa}, of X-ray Binaries \citep{Giacconi:1962aa}, quasars \citep{Schmidt:1963aa}, pulsars \citep{Hewish:1968aa} and Gamma Ray Bursts \citep{Klebesadel:1973aa} in the 1960s, magnetars in the 1970s and Fast Radio Bursts \citep{Lorimer:2007aa} a decade ago have revealed the nonthermal universe. This is a very different world from the thermal universe studied by stellar astronomers. Roughly ten percent of the total gravitational energy that is released in the universe--- principally by black holes and neutron stars ---  (and nearly one percent of the nuclear energy) is carried off or radiated by relativistic particles and these have to be  accelerated. The only way to accelerate charges is through the application of electric field. However, it is hard to sustain large enough electrostatic field for this purpose and moving or rapidly changing magnetic field is generally invoked to generate the necessary potential differences. 

While the prodigious power of high energy processes has long been appreciated \citep[e.g.,][]{Burbidge:1956aa}, it is their rapid variability that has drawn more attention in recent years. This takes us into the regime of {\sl extreme astrophysics}. The associated particle acceleration has to be extremely potent and the steady, stochastic energy conversion of traditional Fermi acceleration is not an option. New mechanisms that can operate in a great variety of sites are needed. A common characteristic of many of these cosmic sources is the extraction of rotational and gravitational energy from a {\sl prime mover} by large scale electromagnetic field. In the case of a pulsar or a black hole it is rotational energy. in the case of a magnetar it is ultimately gravitational energy. In the case of an accretion disk, it is both. We are not confident that this is always the case, but it is beginning to look that way. 

Inevitably, this electromagnetic field is accompanied by plasma, at the very least sufficient to supply the space charge and current, and this plasma is ultimately expected to share the electromagnetic energy. Two quite different pathways for this energy conversion are available. The first is {\sl dissipation}. Electromagnetic energy is converted (``prodigally'') to particle energy at a rate ${\bf E}\cdot{\bf j}$ per unit volume, and entropy is created. The second is {\sl plasma acceleration}. The electromagnetic force density $\rho{\bf E}+{\bf j}\times{\bf B}$ increases the bulk kinetic energy of all of the plasma  (``industriously''). This may be very efficient but does not directly produce dramatic particle acceleration and rapid variability. This may happen later, for example when the bulk flow passes through a strong shock front. (Rapidly variation from the solar wind stimulated \citet{Gold:1955aa} to make the radical proposal that collisionless plasma could form essentially discontinuous shock fronts.)  Of course, electromagnetic field must be present at the shock (in order to mediate the  shock in the absence of collisions),  but this is probably independent of the electromagnetic field associated with the prime mover. 

Observations of rapidly variable sources, especially by $\gamma$-ray telescopes, strongly suggest the presence of regions of relatively high electromagnetic energy density which is efficiently dissipated and where electrons and positrons are efficiently accelerated up to their radiation reaction limits on a light crossing timescale. The available electromagnetic energy is seized, ``plutocratically'', by a small minority of particles and not, ``democratically'', by the bulk of the plasma. Additional features, such as bulk relativistic motion, may be present and needed to account for the observations.  We call this general phenomenon {\sl magnetoluminescence} \citep{Blandford:2014aa,Blandford:2015aa} by analogy with the flashes of light seen from collapsing bubbles created by ultrasound --- sonoluminescence \citep{Brenner:2002aa}. (This phenomenon can also be produced naturally by pistol shrimp \citep{Lohse:2001aa}.) Similar mechanisms may have to be invoked to account for the acceleration of cosmic rays with PeV \newtext{($10^{15}\,{\rm eV}$)}, EeV \newtext{($10^{18}\,{\rm eV}$)} and even ZeV \newtext{($10^{21}\,{\rm eV}$)} energy. 

Our goal in this chapter is to complement some of the other chapters in this volume by providing a general description of extreme particle acceleration that brings out the common features present in the many sources where magnetoluminescence is seen. In the rest of the chapter, we first give a quick summary of the challenges brought up by various extreme particle accelerators in \S\ref{sec:accelerator}, then discuss some basic principles of particle acceleration in \S\ref{sec:principle} and global considerations of electromagnetic dissipation in \S\ref{sec: global}. We show relativistic MHD and kinetic modeling of the possible processes underlying magnetoluminescence in \S\ref{sec: MHD} and \S\ref{sec:kinetic}, respectively. We conclude with a discussion of future observational, simulation and experimental prospects in \S\ref{sec:future}.  

\section{Extreme Particle Accelerators}\label{sec:accelerator}
\subsection{Pulsars and their Nebulae}
A striking and relatively recent example of extreme particle acceleration is provided by the $\gamma$-ray flares in the Crab Nebula, best observed by Fermi \citep[e.g.,][]{Buehler:2014aa}. The nebula itself, which has to be our best laboratory for high energy astrophysics, has been observed over the entire electromagnetic spectrum from $\lesssim100\, {\rm MHz}$ to $\gtrsim1.5\, {\rm TeV}$ \citep[see, e.g.,][for a review]{Hester:2008aa}. It has been powered for the past 963\,y. by the central pulsar with its $\sim30\,{\rm Hz}$ rotation frequency.  The pulsar loses energy at a rate $\sim5\times10^{31}\,{\rm W}$, roughly four times the current bolometric power ($\sim1.3\times10^{31}\,{\rm W}$) of the nebula, which peaks around $\sim10\,{\rm eV}$. The power assuredly takes the form of an electromagnetic Poynting flux leaving the surface of the neutron star, becoming a relativistic electromagnetic wind with properties that are best defined by the ratio of the electromagnetic energy flux to the particle energy flux (usually assumed to comprise pairs, though ion-dominated winds are possible), $\sigma$, the outflow Lorentz factor and a description of the variation of the outflow with latitude. The Crab Nebula demonstrates that electromagnetic conversion of prime-mover power into relativistic electrons and positrons can be very efficient. However, it has not told us where this conversion takes place and there are viable models that locate this conversion all the way from the pulsar light cylinder to the outer nebula \citep[e.g.,][]{Kirk:2009aa, Arons:2012aa}.

The Crab pulsar wind must decelerate from the speed of light to the expansion speed of the supernova remnant $\sim1500\,{\rm km\,s}^{-1}$. This is commonly supposed to begin at a strong, relativistic, collisionless shock front located where the outflow momentum flux matches the ambient nebula pressure at a radius \newtext{$\sim3\times10^{15}\,{\rm m}\sim0.1\,{\rm pc}$} \citep{Rees:1974aa, Kennel:1984aa}. If, as expected, the strength of the wind decreases significantly with latitude, the shock surface will be quite oblate \citep[e.g.,][]{Komissarov:2003aa}.  This is compatible with the observation of an equatorial torus at X-ray energies \citep{Weisskopf:2000aa}. 

However, there is no observation that suggests the presence of a shock outside the equatorial zone. The ``inner knot'', a compact feature located $0.65''$ southeast of the Crab pulsar in infrared and optical images \citep{Hester:1995aa,Rudy:2015aa}, has been modeled as a point on the oblique part of the termination shock in the polar region\citep{Komissarov:2004aa,Komissarov:2011aa}, but there is some evidence that it may be associated with the low latitude section of the pulsar wind where the effective magnetization becomes low because of the alternating magnetic flux in a ``striped'' wind \citep{Lyutikov:2016ab,Yuan:2015aa}. At higher latitudes, it is consistent with $\sigma\gg1$ at this radius (as we must assume for the mechanisms we are reviewing here) so that the shock is weak --- there may be several of them --- and most of the deceleration happens gradually. Given this model, the magnetic field strength is $\sim100\,{\rm nT}$ here, decreasing to $\sim30\,{\rm nT}$ in the body of the nebula.

The observed $\gamma$-ray flares \citep[e.g.,][]{Buehler:2014aa} happen with a cadence $\sim1\,{\rm y}$ and have only been seen at energies $\sim300\,{\rm MeV}$ and there are no associated pulsar timing glitches. Variation, on timescales as short as a few h has been reported. The peak isotropic luminosity is roughly $10^{29}\,{\rm W}$ and the energy radiated is $\sim10^{34}\,{\rm J}$. The flares and secular observations \citep{Wilson-Hodge:2011aa} demonstrate that the energy conversion is intermittent not steady and that the mechanism can be locally cataclysmic.


\subsection{Blazars}
Most galaxies, at least as massive as our Galaxy, possess nuclei surrounding a massive  ($\sim10^6-10^{10}\,{\rm M}_\odot$) black hole which become {\sl active} either when they are supplied with gas that is efficiently accreted or the holes are spun up and the rotational energy is tapped \citep[e.g.,][]{meier_black_2012}. A common expression of this activity is the formation of a pair of antiparallel {\sl relativistic jets}, with Lorentz factors $\Gamma\sim10$, presumably directed along the hole's rotation axis, at least initially. The presence of these jets was originally inferred from the observations of large double radio sources straddling the optical image of the host galaxy, which they supply with energy, momentum and mass. However, jets are now observed throughout the electromagnetic spectrum, most notably at $\gamma$-ray energies \citep[see, e.g.,][for a review]{Madejski:2016aa}. As the outflows are relativistic close enough to the hole, jet emission is strongly beamed and a small minority of sources are observed to be unusually bright as they are directed towards us. These are called {\sl blazars} and they can dominate a flux-limited sample. Observations made using Very Long Baseline Interferometry (VLBI) at cm and mm wavelengths have confirmed that these jets are already collimated on scales $\lesssim100-1000$ black hole gravitational radii \citep{Marscher:2008aa} and we are now on the threshold of resolving the curved spacetime around the black hole \citep{Broderick:2015aa,Fish:2016aa}. Relativistic jets are now acknowledged as being powered by the inner accretion disk and the black hole spin energy but the relative importance of these two sources is debated. 

Blazars are frequently divided into two types, {\sl BL Lac objects}, or BLL, and {\sl Flat Spectrum Radio Quasars} or FSRQ. BLL are low total power \newtext{objects} and mostly local; FSRQ are high power quasars which can outshine the host galaxy and are mostly at cosmological redshifts. Both classes exhibit remarkable GeV/TeV flares with source variation times as short as $\sim2-3$ minutes \citep[e.g.,][]{Albert:2007aa,Aharonian:2007aa,Aleksic:2011aa,Ackermann:2016aa}. This suggests that the emission originates very close to the jet source. However, there is a lower bound on the radius where the particle acceleration and emission can occur due to absorption involving pair production on soft photons. This radius can be calculated assuming QED (in which we should have absolute confidence!) and this defines the {\sl gammasphere} \citep{Blandford:1995aa}. As with the Crab Nebula, there are viable models with emission radii ranging from the gammasphere to the sites of recollimation located $\gtrsim100\,{\rm pc}$ from the hole. There is now good evidence that the $\gamma$-rays are produced before and within the radio emission \citep{Max-Moerbeck:2014aa}. 

\subsection{Gamma Ray Bursts}
Most {\sl Gamma Ray Bursts}, GRB, are cosmologically distant and release the energy equivalent of a significant fraction of a stellar mass \citep[see, e.g.,][for a review]{kouveliotou_gamma-ray_2012}. There are also two types\newtext{: the}  {\sl short bursts} lasting $\lesssim2\,{\rm s}$ are conjectured to be neutron star binaries, merging under gravitational radiation reaction and potentially observable by LIGO/VIRGO\newtext{; the} {\sl long bursts} are convincingly associated with a subset of core-collapse supernovae tentatively identified with progenitors that have lost their helium and hydrogen envelopes although other factors could easily be relevant. GRBs are also relativistic jets  (with $\Gamma\sim100-1000$). They are collimated by the collapsing star in the long bursts; the jets from short bursts may be collimated by a very dense accretion disk.

The prompt $\gamma$-ray emission from GRBs is often found to be variable on timescales as short as $\sim10\,{\rm ms}$.  The {\sl afterglow}, observed throughout the electromagnetic spectrum, is formed later when the jets are decelerated by interaction with the circumstellar medium.

One common feature of models of long and short bursts and AGN (and, also, Galactic superluminal sources \citep{Mirabel:1999aa}) is the realization that the accreting gas near the hole cannot cool requiring the formation of a thick orbiting torus at small radius which defines two funnels that may define the initial jet shape \citep{Yuan:2014aa}. (Jets are also formed by pulsars like the Crab pulsar, \newtext{and here the hoop stress of the toroidal magnetic field downstream of the termination shock may be responsible for the confinement \citep{Komissarov:2003aa}.}) There are actually three accretion regimes. At low mass supply rates the electrons may be much cooler than the ions and unable to cool the gas on an inflow timescale. Conversely, at a high mass supply rate, the gas can radiate but the Thomson opacity is so high that the photons are convected inward faster than they can escape outward. It is only at an intermediate mass supply rate that a thin accretion disk will form. On quite general grounds, the low and high supply regimes require powerful winds to carry off the energy that is released during accretion \citep{Blandford:1999aa}.  

\subsection{Magnetars}
A minority of neutron stars, called {\sl magnetars} have surface field strengths in the 10-100 GT range over two orders of magnitude more than typical radio or X-ray pulsars \citep[see][for a review]{Kaspi:2017aa}. They are found to be rotating with periods $\sim1-10\,{\rm s}$ range but could have been borne with millisecond periods and, if so, they would be strong candidates for the prime-movers of GRBs. Magnetars show dramatic flares \citep{Evans:1980aa,Hurley:1999aa,Hurley:2005aa,Mereghetti:2005aa,Boggs:2007aa} which are likely to be caused by magnetic instability associated with either the crust or the magnetosphere \citep{Thompson:1995aa,Thompson:1996aa}. (The energy release is far too much to be derived from rotation.) The emission sites are likely to be electromagnetically-dominated. The underlying electromagnetic instability may involve rapid, relativistic reconnection of flux tubes, or may be simple untangling with small changes of magnetic helicity. Shocks and non-relativistic reconnection may account for the acceleration needed for more slowly varying emission from these sources.

\section{General Principles}\label{sec:principle}
\subsection{Steady vs Impulsive Acceleration}
We have emphasized the rapid variability in these sources because it provides an important clue as to the underlying physical conditions \citep{Begelman:2008aa}. However, much of the power is associated with slowly varying source components and extreme particle acceleration is not required. Nonetheless, it is still the case that the radiative efficiency must be high which implies that the associated cooling timescale should be short compared with the expansion timescale in an outflow and the timescales for other non-radiative losses. 

The most commonly invoked means for converting large scale electromagnetic energy into high energy particles and radiation is through magnetic reconnection \citep{biskamp_magnetic_2000,Priest:2000aa}. There has been much effort devoted to understanding nonrelativistic reconnection especially in the context of space and solar physics and also for Tokamaks. The basic idea is that magnetic field is convected by the flow, more or less incompressibly with little dissipation almost everywhere. However, small regions are created where the field reverses abruptly and the current density increases until sufficient dissipation is produced to allow the magnetic field lines to ``exchange partners''\newtext{, or ``reconnect'',} with a loss of magnetic energy. It is generally the case that surprisingly energetic particles are produced but with low efficiency. Reconnection need not be steady; various instabilities, like the {\sl tearing mode}, can occur \citep[e.g.,][]{Furth:1963aa}. Also the microscopic description requires the tensor character of the electrical conductivity to be included \citep[e.g.,][]{Wang:2000aa}. In \S\ref{subsec: reconnection}, we will discuss some basic particle acceleration mechanisms in reconnection.

Relativistic reconnection is now being simulated and is clearly much more promising as an agency of extreme particle acceleration \citep[e.g.][]{Zenitani:2001aa,Guo:2014aa,Sironi:2014aa,Werner:2016aa}. However, non-relativistic reconnection may be a slow process, and as such would be better suited to steady acceleration to modest energy, rather than the extreme, ``impulsive'' acceleration needed for the the most dramatic flares.

\subsection{Basic Particle Acceleration in Reconnection} \label{subsec: reconnection}

\begin{figure}[htb]
  \centering
        \includegraphics[width=\textwidth]{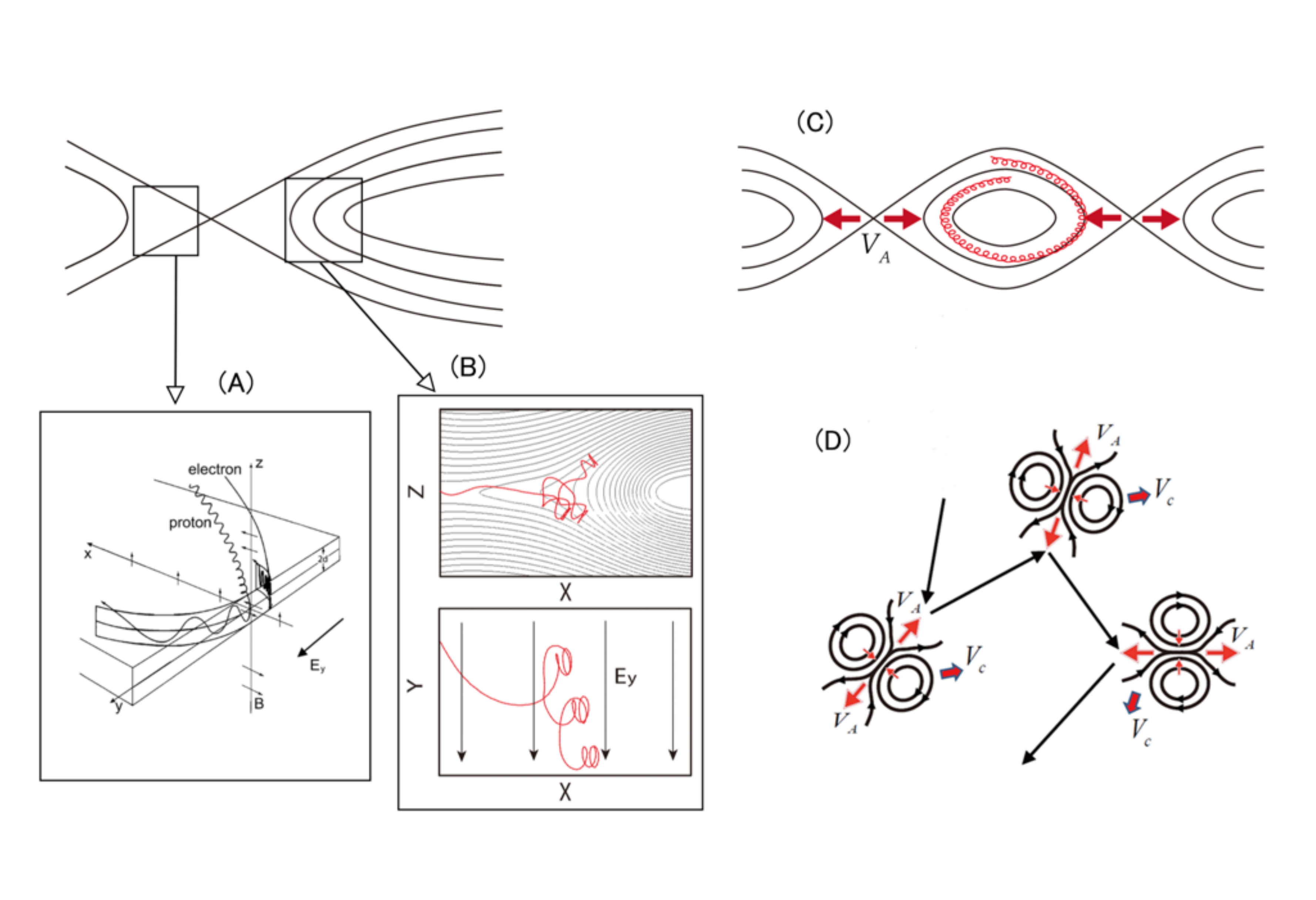}
  \caption{Typical particle orbits accelerated during magnetic reconnection.  (A) {\sl Speiser} motion/meandering motion, adapted from \citep{Speiser:1965aa}. Particles can get accelerated during the meandering/bouncing motion along the electric field, under the anti-parallel magnetic field $B_x$, (B) the gradient/curvature $B$ drift motion in the magnetic field pileup region, (C) the trapped particles in the shrinking magnetic islands, (D) the stochastic Fermi acceleration in multiple magnetic islands.}\label{fig:AcceleratingParticles}
\end{figure}

There has been much simulation study of the particle acceleration in magnetic reconnection, and several different acceleration processes operating in \newtext{various regions} in the multi-scale reconnection system have been proposed.  Since the dominant electric field responsible for particle acceleration is the inductive electric field during time evolution of reconnection, the magnetic diffusion region, which coincides with the so-called X-type region, would be the principal particle acceleration site.  As shown in Figure \ref{fig:AcceleratingParticles} (A), the charged particles, whose motions are described by {\sl Speiser} (meandering) motion \citep{Speiser:1965aa}, can almost directly resonate with the inductive electric field and can be quickly accelerated in the diffusion region where the magnetic field is weak.  This acceleration may be almost free from synchrotron radiation loss.  However, the volume size of the X-type region is small for the standard {\sl Petschek} reconnection model associated with a pair of slow mode shocks \citep{Petschek:1964aa}, and the total energy release rate would not be necessarily large.  For the case of {\sl Sweet-Parker} reconnection with an elongated diffusion region in plasma sheet \citep{Parker:1963aa}, there may be enough volume for direct acceleration in the meandering motion by the electric field, but the magnitude of the electric field is weak compared to Petschek reconnection, because the dimensionless reconnection rate in the Sweet-Parker model is estimated as $1/\sqrt{R_m}$, which should be contrasted with the reconnection rate in Petschek model $\sim1/\ln(R_m)$, where $R_m=L V_A/\eta$ is the magnetic Reynolds number based on the inflow Alfv{\'e}n velocity $V_A$, which is also frequently referred to as Lundquist number, $L$ is the length of the reconnection region and $\eta$ is the magnetic diffusivity.  

In addition to the {\sl Speiser} motion in the diffusion region, particles can be accelerated in the pileup magnetic field region produced by compression of the reconnection outflow. The collision of the reconnection outflow emanating from the diffusion region with the pre-existing plasma at rest will produce the pileup region associated with a large jump of magnetic field.  Figure \ref{fig:AcceleratingParticles} (B) shows that the $\nabla B$ and/or curvature $B$ drift motion for ions/electrons, which is parallel/anti-parallel to the inductive electric field, contributes to the energization \citep{Hoshino:2001aa}.  If the particles are magnetized and their gyro-radii are small compared with the scale height of the pileup magnetic field, the first adiabatic invariant $p_{\perp}^2/B$ is conserved, and a large energy gain can be expected due to the large jump of the magnetic field intensity. 

Not only the X-type and the B pileup regions in the single reconnection site, but magnetic islands surrounded by the loop-like magnetic field lines, which situation can happen for multiple reconnection/tearing mode instability in a long current sheet, can invoke particle energization, shown in Figure \ref{fig:AcceleratingParticles} (C).  Because of the shrinking magnetic field lines bounded by two X-type regions at each end, the particles trapped by the magnetic island can gain energy by reflection from both ends of the contracting magnetic field line \citep{Drake:2006aa}.  The energy gain in this Fermi acceleration process can be estimated from the conservation of the second adiabatic invariant $p_{\parallel} L$, where $L$ is the loop length of the magnetic island.  However, as being energized by the contracting magnetic field line, the particles will escape from the island if their gyro radii become comparable with the size of the island and the particle acceleration will cease.

However, it is possible that high energy particles can be further accelerated in a much larger plasma environment with many magnetic islands. By analogy to the original Fermi acceleration model, where particles gain energy stochastically during head-on and head-tail collisions of particles with many magnetic clouds, let us replace the magnetic clouds with magnetic islands, as shown in Figure \ref{fig:AcceleratingParticles} (D).  In the original Fermi acceleration, the increase in particle energy is known to be \newtext{of} second order in $V_c/c$, where $V_c$ is the speed of random motion of magnetic clouds. However, if there are many magnetic islands with the Alfv\'{e}nic plasma outflow, it is suggested that the particles can preferentially interact with the reconnection outflow region, because the energetic particles have a tendency to be situated outside the magnetic island during acceleration.  As a result, the increase in energy becomes first order of $V_A/c$, where $V_A$ is the Alfv\'{e}n speed \citep{Hoshino:2012aa}.

Regardless of the acceleration mechanisms in details, another key issue for a rapid magnetic energy dissipation and particle acceleration is how the large electric field is generated during reconnection.  The average electric field in the reconnection system is basically determined by $V \times B/c$ induced by the inflow and outflow motions, and roughly speaking the outflow speed can \newtext{reach the} Alfv\'{e}n speed for both non-relativistic and relativistic reconnections. Since the Alfv\'{e}n speed is given by $V_A = c \sqrt{\sigma/ (\sigma+1)}$,  the fast reconnection flow and the large electric field can be realized in a strong magnetized plasma with a large $\sigma$ value.  The spontaneous vs driven reconnection is another key issue to control the reconnection rate.  It is known that small but finite Poynting flux injection from the inflow boundary into the system can drive a fast reconnection for both linear and nonlinear stages.  While the linear growth rate for the spontaneous tearing mode is proportional to $R_m^{-3/5}$ \citep{Furth:1963aa}, the growth rate for the driven case is $R_m^{-1/3}$ \citep{Horton:1988aa}.  The enhancement of the reconnection rate can be also found in the double tearing mode, which is characterized by the presence of two (or more) tearing current sheets whose surfaces are attached to each other.  The reason of enhancement of reconnection rate is simply because the surface can share the inflow and outflow boundary condition for two tearing islands, which is the same situation as arises in driven reconnection.

\subsection{Pairs vs Hadrons}
There has been much discussion of the nature of the positive charge carriers especially in jets. Are they positrons or protons? Given the large EMFs generated, the former are readily created through $\gamma$--$\gamma$ or even $\gamma$-$B$ processes although this may be balanced by pair annihilation at the base of the jet. However hadrons are likely to be incorporated into the flow through entrainment from the surrounding medium, which might comprise gas accreting at high latitude, gas clouds in the interstellar medium, a thick, confining torus or an outflowing wind. Probably all four possibilities are relevant. It may be that this is the explanation for the famous \citet{Fanaroff:1974aa} dichotomy of double radio sources, with the low power, Type 1 sources being identified with jets that are efficiently decelerated by their surrounding and Type 2 sources containing jets that are not slowed down. If the proton density is high, shocks are likely to be very important particle accelerators  but conditions will then not be propitious for extreme acceleration/magnetoluminescence.

\subsection{Emission Mechanisms}
Although we surely trust the physics of the underlying radiative processes, we are not completely confident that we understand which of them operates and where. The ``Bactrian'' (two-humped) spectrum characteristic of blazars has been commonly interpreted as a two process --- synchrotron and Compton --- emission from a single, homogeneous source. However, jets are observed to radiate all along their lengths and it seems more likely that both humps comprise emission from different radii. 
Interestingly, there is a maximum synchrotron photon energy that can be radiated in a simple electromagnetic source. If we balance the rate of energy gain $\sim e Ec$ by the radiative loss and impose the condition $E\lesssim B$, then the maximum photon energy is found to be $\sim m_ec^2/\alpha$, where $\alpha$ is the fine structure constant. This is $\sim70\,{\rm MeV}$. 

It is common to distinguish two types of Compton radiation with externally produced photons dominating in the FSRQs and internal synchrotron photons dominating in the BLLs. Again, it seems more reasonable to suppose that both options are exercised along the jet. 

The self-absorbed  synchrotron radio emission seen from jets, known as the {\sl core}, sometimes seems to vary so rapidly that the emission mechanism may have to be coherent, perhaps a cyclotron maser \citep{Begelman:2005aa}. An important constraint is that the radio waves have to avoid absorption due to the induced Compton effect or stimulated Raman scattering which sets an upper limit on the electron column density along the line of sight \citep{Levinson:1995aa}.

There are more options if hadrons are present. There have been several suggestions that the $\gamma$-rays are produced as a consequence of pion production following proton-proton collisions. The neutral pions decay efficiently into gamma rays. However, more of the energy goes into charged pions where the decays involve muon and electron neutrinos and may produce particles that are insufficiently energetic to radiate efficiently. A variation which deserves more attention is for a proton be accelerated to very high energy so that it can photo-produce electron-positron pairs which quickly radiate by the synchrotron process into the ``Compton''  hump at energies above the conventional synchrotron limit.  This might be important in a high radiation density environment in a GRB or a quasar. 
\subsection{Relativistic Beaming}
Doppler beaming is an essential feature of electromagnetic flows as the characteristic signal speed is necessarily relativistic.  The easiest way to handle this is to introduce a {\sl Doppler factor} ${\cal D}=[\Gamma(1-{\bf V}\cdot{\bf n}/c)]^{-1}$, where $\Gamma=(1-V^2/c^2)^{-1/2}$, for Lorentz transformation from a (primed) frame in which we can evaluate the emissivity $j'_{\nu'\Omega'}(\nu',{\bf n}')$ and the absorption coefficient $\kappa'(\nu',{\bf n}')$. $\bf V$ is the velocity of the primed frame in the frame of a distant, though not cosmologically distant, observer along a direction $\bf n$. Primed frame quantites are transformed into observer frame quantities according to $dt={\cal D}^{-1}dt'$, $\nu={\cal D}\nu'$, ${\bf n}'={\cal D}{\bf n}-\Gamma({\cal D}+1){\bf n}/(\Gamma+1)$,  $j_{\nu\Omega}(\nu,{\bf n})={\cal D}^2j_{\nu'\Omega'}(\nu',{\bf n}')$, $\kappa_{\nu\Omega}(\nu,{\bf n})={\cal D}^{-1}\kappa_{\nu'\Omega'}(\nu',{\bf n}')$ \citep[e.g.,][]{Blandford:1979aa}. The intensity, which transforms according to $I_{\nu\Omega}(\nu,{\bf n})={\cal D}^3I_{\nu'\Omega'}(\nu',{\bf n}')$, is best considered in the observer frame and can be corrected for cosmological expansion if appropriate. For a source that is beamed towards us, it is a good start to approximate these formulae by setting ${\cal D}\sim\Gamma$.

With the increasing sophistication of simulations of electromagnetic sources, it is worth carrying out more careful radiative transfer calculations. These are best prosecuted in the unprimed, observer frame, transforming the emissivity and absorption coefficient into this frame.
\section{Global Issues}\label{sec: global}
\subsection{Voltages and Currents}
Before discussing details, it is helpful to consider the global energy balance in a relativistic, electromagnetic source. In the most simple-minded \newtext{model} of a spinning conductor with angular frequency $\Omega$ and threaded by magnetic flux $\Phi$, the potential difference generated is $V_{\rm eff}\sim\Omega\Phi/2\pi$. There will be a current flow $I_{\rm eff}$  outside the body and the effective impedance will be roughly that of free space $Z_{\rm eff}\sim100\,\Omega$ so that $I_{\rm eff}\sim V_{\rm eff}/Z_{\rm eff}$.  The power that can be effectively dissipated as particle acceleration is then $L_{\rm eff}\sim V_{\rm eff}I_{\rm eff}\sim\Omega^2\Phi^2/4\pi^2Z_{\rm eff}$. In the case of a magnetar where the energy release is gravitational, $\Omega$ should be replacedby $\Omega_{\rm eff}$ the reciprocal of the characteristic timescale associated with the release of seismic energy.

Let us give some examples. For the Crab pulsar, \newtext{$V_{\rm eff}\sim50\,{\rm PV}\sim 5\times10^{16}\,{\rm V}$} and $L_{\rm eff}\sim3\times10^{31}\,{\rm W}$, comparable with the bolometric luminosity of the Crab Nebula. For a quasar like 3C279, \newtext{$V_{\rm eff}\sim300\,{\rm EV}\sim3\times10^{20}\,{\rm V}$} and $L_{\rm eff}\sim10^{39}\,{\rm W}$ and for a powerful GRB, \newtext{$V_{\rm eff}\sim100\,{\rm ZV}\sim10^{23}\,{\rm V}$} and $L_{\rm eff}\sim10^{44}\,{\rm W}$. Note that there is sufficient potential difference to accelerate Ultra High Energy Cosmic Rays (UHECR) in powerful AGN though whether or not these particles can escape without suffering catastrophic photopion production loss is debatable. Newly born magnetars, with $10^{11}\,{\rm T}$ surface magnetic field and an initial spin rate close to the centrifugal breakup limit, could generate an EMF \newtext{$V_{\rm eff}\sim10\,{\rm ZV}\sim10^{22}\,{\rm V}$}; they, too, have been considered as viable accelerators for UHECR \citep{Arons:2003aa}.

A very simple effective circuit description raises the question of where do these currents close. If this is closeby, the flow becomes gas dynamical and remote acceleration is likely to be associated with shocks and shear flows.  Alternatively, the current  may persist to the observed extremity of the source, for example the hot spots associated with a powerful double radio source. In this case, the processes discussed here are likely to be relevant to most of the emission.

\subsection{Flow of Flux}
The voltage $V_{\rm eff}$, measured in V is also the flow of magnetic flux measured in Wb s$^{-1}$. If as is the case with a pulsar wind nebula, the flux is confined within a volume that is expanding much slower than the speed of light then the flux must be dissipated at almost the rate at which is supplied. Particle acceleration in one form or another is inevitable. (If this did not happen and the walls were perfect conductors, then the electromagnetic field would react back on the source.) The only questions are where does it occur and what are the resulting particle energy distributions. For an axisymmetric rotating source, the two most promising possibilities are the symmetry axis where the polar current will be concentrated and along the path of the return current, specifically along the equatorial  plane in a pulsar wind nebula and along the jet walls in the case of AGN jets. There is observational evidence for both of these \newtext{\citep{Hester:1995aa,Weisskopf:2000aa,Reid:1989aa}}. 
\section{Relativistic MHD Description}\label{sec: MHD}
\subsection{Knots and Tangles}
Although there may be some validity to describing the electromagnetic field under the force-free approximation close to the source, the inertia of the plasma is likely to become significant quite quickly and relativistic MHD will be the preferred fluid description although we have little understanding of how the particle density builds up and is regulated. Nonetheless it is reasonable to expect that MHD instabilities will break any axisymmetry and lead to the formation of current sheets. As is observed in the solar corona, ``hairy'' magnetic ropes are likely to form, twisted by field-parallel (zero-stress) current. The extent to which this is a fair description of these magnetically dominated sources is unknown but if we adopt it, it is easy to believe that the writhing {\sl magnetic ropes} can become both tangled and knotted. 

\begin{figure}[htb]
  \centering
        \includegraphics[width=0.7\textwidth]{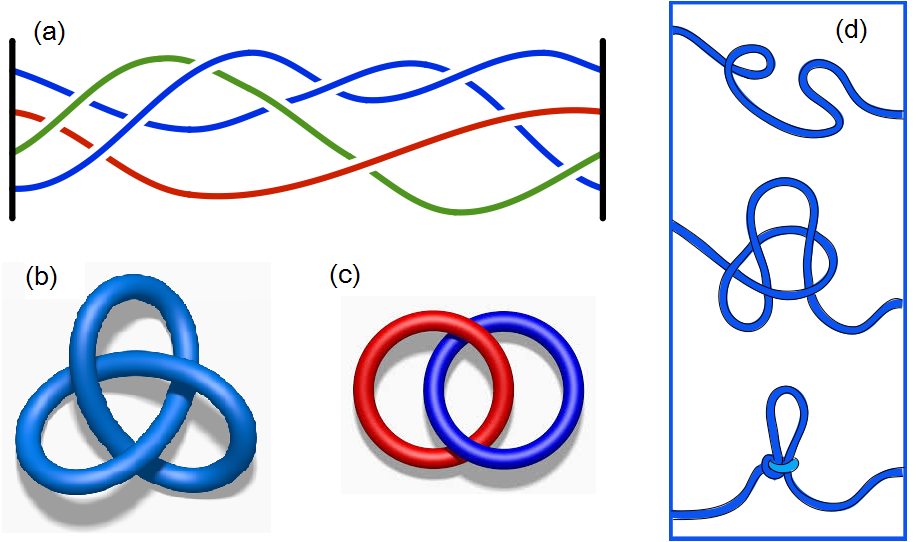}
  \caption{An illustration of topologies the magnetic ``ropes'' may take. (a) Braid. (b) Knot. (c) Link. (d) Hitch, which is topologically equivalent to a straight line.}\label{fig:knots}
\end{figure}

The distinction is important. A knot, which contributes to the magnetic helicity $H=\int dV {\bf A}\cdot{\bf B}$ \citep{moffatt_magnetic_1978,Bellan:2000aa}, is topologically distinct from a tangle or ``unknot'' (Fig. \ref{fig:knots}). If the magnetic field that is created by the prime mover is required to evolve according to the precepts of perfect MHD, it will remain unknotted. Even if $H$ were non-zero initially, it would decrease in inverse proportion to the scale size of the flow. However, if magnetic ropes form, it is an inevitable consequence of instability in these outflows that there will be places where they will be pulled or pressed against one another. Reconnection of pairs of ropes at unrelated locations seems to be inevitable with an overall release of energy through relaxation of magnetic tension and transverse expansion. The knots that are then created topologically will be initially quite loose but are likely to tighten under the action of magnetic tension with the passage of time. Additional instances of forced reconnection  will resolve these knots, leading to further energy release and particle acceleration. (Knots are very important in the history of physics as their {\sl aetherial}  expression was once the most popular model of atoms \citep{Tait:1907aa}. They have also played a prominent role in modern field theory.) 

However, it also seems to be quite likely that tangles of individual ropes will develop and tighten into {\sl slip knots} or {\sl hitches} that may be resolved without large-scale reconnection simply through an increase in magnetic tension. As with conventional ropes the ease with which this can happen depends upon the surface friction. In the case of magnetic ropes this will depend upon small-scale electromagnetic stresses associated with the ``hair''\newtext{---wandering field lines near the boundary of the flux ropes}. If the friction is large, then large tangles with large latent energy can be created. 

From a particle acceleration standpoint, tangles can untangle at effectively the speed of light in a relativistic source and can be responsible for the flares; knots require reconnection, which may be slower to resolve, and may be responsible for the higher power, steady emission.

\subsection{Sheets, Shocks and Caustics}
Many of the relativistic sources that we have been discussing, such as blazar jets, are observed to be expanding relativistically. Others, such as pulsar wind nebulae, have relativistic sound speeds and can therefore sustain internal relativistic motions. Doppler boosting with Lorentz factor $\Gamma$, as summarized in Sec. 3.4, is therefore likely to be present and to contribute to the observed flux. If we idealize the source as a sphere moving with a single velocity, then we may see variation as the source expands in its own reference frame. Alternatively, the velocity may turn through an angle $\gtrsim\Gamma^{-1}$ and this will produce a roughly symmetric pulse. Parallel acceleration should produce an asymmetric pulse. Changes in the optical depth at either radio or $\gamma$-ray energy can also produce rapid variation.

However, there are additional possibilities if the emission comes from a surface, such as a current sheet or a shock front that is moving ultrarelativistically. An element of the surface may beam the synchrotron or Compton radiation in a narrow cone about its direction of motion. (Among several factors which could complicate a detailed model, the speed of the emitting surface might differ from the speed of the emitting plasma, as happens in a shock front. In addition the electrons and positrons might be carrying a current and moving with different speeds.) Now, for a generic, corrugated surface, the mapping from the moving surface onto a distant sphere might be many to one which can lead to the formation of caustics \citep[cf,][]{Lyutikov:2012aa}. The simplest generic caustic is a {\sl fold} which produces an asymmetric flux variation $\propto|t-t_0|^{-1/2}$ before or after the time $t_0$ when a pair of images either disappears or appears. Effects such as these could enhance the observed variability from sources where extreme particle acceleration might be taking place.

\subsection{MHD Simulations}\label{subsec:MHD}
In recent years much progress has been made in relativistic MHD simulations to address the plasma dynamics in pulsar wind nebulae, the formation, propagation and acceleration of jets from black holes/neutron stars, the propagation of shocks/blast waves in GRBs, and so on. We are gradually gaining a reasonable picture of how the magnetized outflow from the engine shapes its environment, as described in many other chapters of this book. Meanwhile, a lot more work is needed to understand the actual dissipation process. Here, as a good starting point, we can focus on a local region in the plasma outflow that has a tangled magnetic field topology, to see whether a rapid conversion of electromagnetic energy to particle kinetic energy can occur.

As one example, \citet{East:2015aa} considered a family of force-free equilibria in a 3D periodic Cartesian box. These configurations have the topology of multiple flux ropes packed in a complex manner. They are solutions to the force-free equation $\nabla\times\pmb{B}=\lambda\pmb{B}$, where $\lambda$ is a constant throughout the space, so each of them has a single characteristic wavelength \citep{Chandrasekhar:1957aa,Moffatt:1986aa,Rosenbluth:1979aa,Bellan:2000aa}:
\begin{equation}\label{eq:Cartesian_periodic_B}
\pmb{B}=\sum_{|\pmb{k}|=|\lambda|}\hat{n}_{\pmb{k}}\times \nabla \chi_{\pmb{k}} +\frac{1}{\lambda }\nabla \times \left(\hat{n}_{\pmb{k}}\times \nabla \chi_{\pmb{k}} \right)
\end{equation}
Here $\chi_{\pmb{k}}$ is the solution to the scalar Helmholtz equation $\nabla ^2\chi +\lambda ^2\chi =0$, $|\pmb{k}|=\lambda$, and $\hat{n}_{\pmb{k}}$ is a constant unit vector. At a fixed total helicity $H$, if a configuration has a shorter wavelength, namely, it contains a larger number of smaller flux ropes, then its total magnetic energy is higher. 

Constant-$\lambda$ force-free configurations have historical importance, because Taylor's conjecture  suggests that a closed, magnetically dominated plasma tends to relax to lower energies while conserving the total helicity, and the relaxed state is just one such constant-$\lambda$ configuration \citep{Taylor:1974aa,Taylor:1986aa}. 

MHD/force-free simulations by \citet{East:2015aa} found that, typically, the shorter wavelength equilibria are unstable to ideal MHD modes; the instability grows on Alfv\'{e}n crossing time scales, and in the nonlinear regime the system goes through a turbulent relaxation process, dissipating a finite amount of energy within just one dynamic time scale. Eventually the plasma settles into the longest possible wavelength. 

\begin{figure}[htb]
  \centering
        \includegraphics[width=\textwidth]{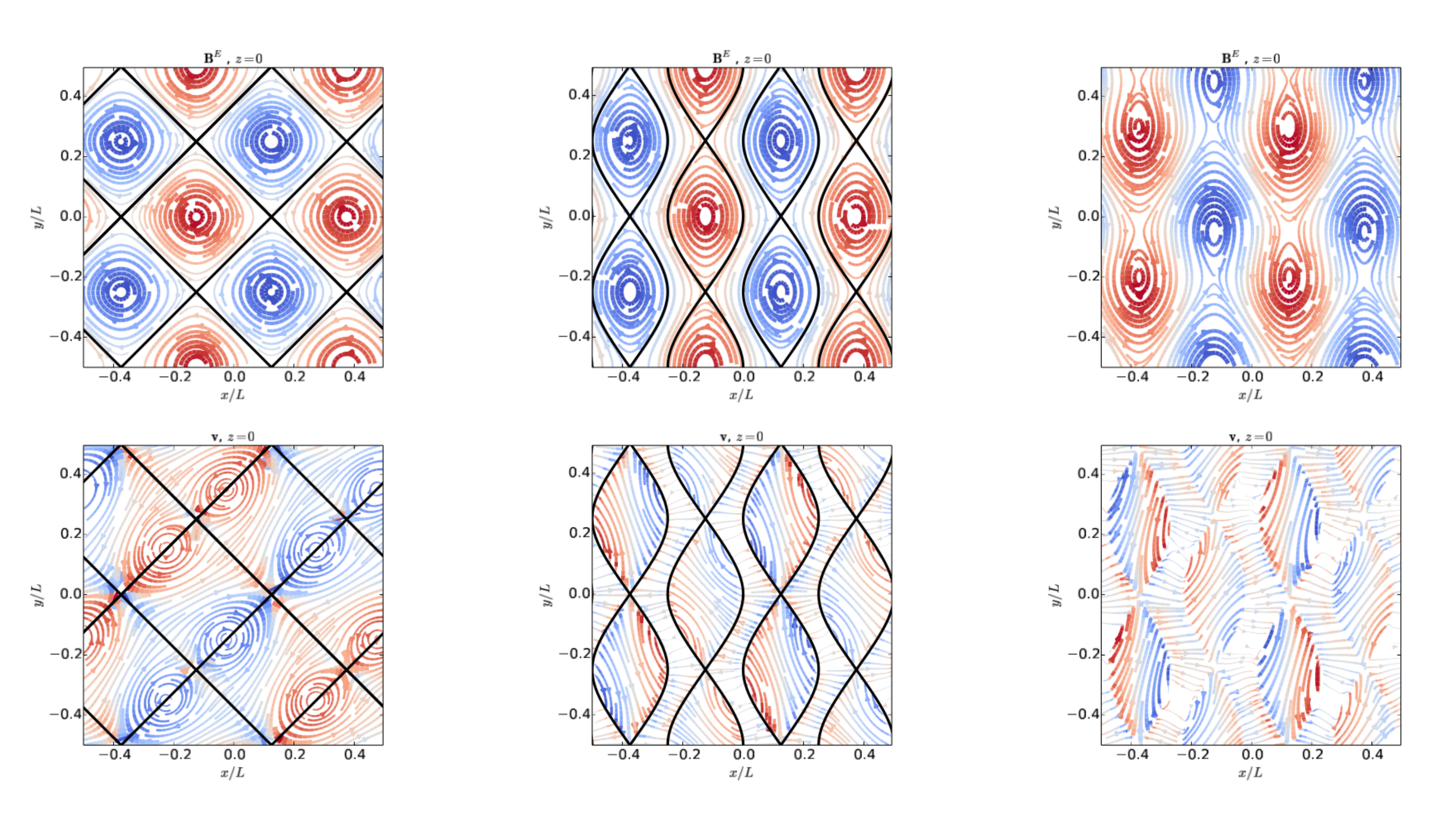}
  \caption{Streamlines of the magnetic field in the initial force-free equilibria for three different examples (top row), and the corresponding velocity field $\pmb{v}=\pmb{E}\times\pmb{B}/B^2$ of the maximally growing unstable mode found from the force-free simulations (bottom row) \citep[from][]{East:2015aa}. The first two equilibria are 2D while the third one is 3D. All plots show streamlines on the $z=0$ plane. The color indicates the perpendicular vector component with red and blue representing, respectively, out of the page and into the page. The thickness of the streamline is proportional to the vector magnitude. The black lines indicate the location of the separatrices in the equilibrium solutions.}\label{fig:Cartesian_simulation}
\end{figure}

\begin{figure}[htb]
  \centering
        \includegraphics[width=0.7\textwidth]{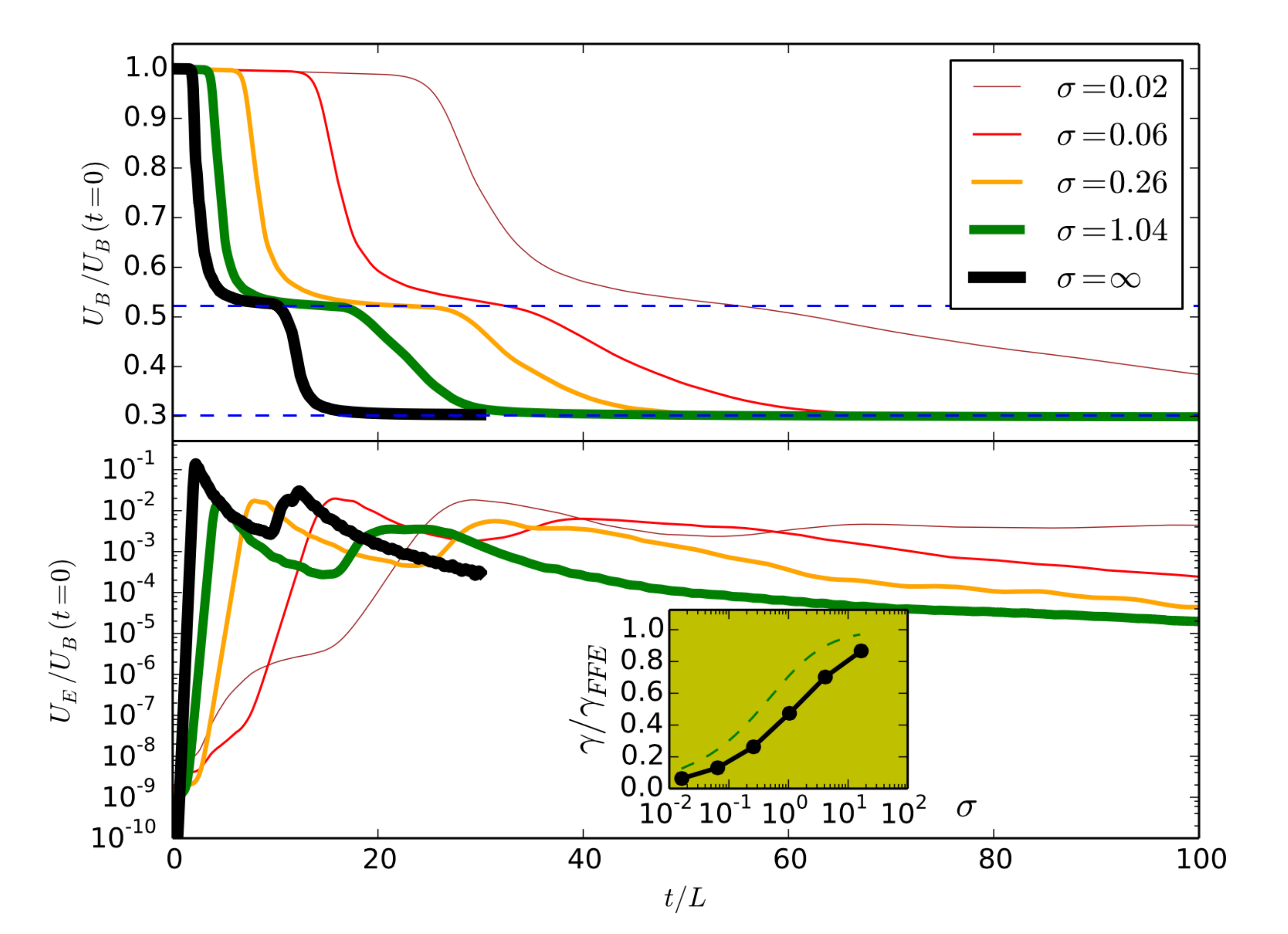}
  \caption{A comparison of the decay of an $\lambda^2 = 11$ equilibrium in simulations with different values of volume averaged magnetization $\sigma\equiv\langle B^2/(4\pi w)\rangle$, where $w$ is the plasma enthalpy \citep[from][]{East:2015aa}. Shown is the magnetic energy (top) and kinetic/electric field energy (bottom). The horizontal dashed lines in the top panel indicate the magnetic energy of $\lambda^2=3$ and $\lambda^2=1$ states with the same helicity. The bottom inset shows the linear growth rate $\gamma$ measured for runs having different magnetization parameters, along with the Alfv\'{e}n speed (dashed line) for comparison.}\label{fig:FFMHD}
\end{figure}

Figure \ref{fig:Cartesian_simulation} shows a few examples of the force-free equilibria and the corresponding maximally growing unstable mode. It suggests that the instability can be viewed as shearing and merging of the flux ropes. The velocity field appears to have non-smooth features, reminiscent of spontaneous current sheets that occur at the flux tube boundaries. In Figure \ref{fig:FFMHD}, it is shown that different volume averaged magnetization values $\sigma\equiv\langle B^2/(4\pi w)\rangle$ (where $w$ is the plasma enthalpy) give the same intermediate and final energy levels, consistent with conservation of magnetic helicity. The amount of magnetic energy dissipated during the evolution thus corresponds to the {\sl free} magnetic energy, defined as the energy difference between the initial configuration and the ground state (longest wavelength state), compared at the same helicity. 

These systematic numerical investigations suggest that, in general, the generic short wavelength, constant-$\lambda$ force-free states in 3D periodic boxes are unstable; the instability is characterized by an exponentially growing electric field in the linear phase, and eventually leads to current sheet formation where significant amount of the free magnetic energy is dissipated within just a single light crossing time. 

\citet{Lyutikov:2016aa} also studied independently a subset of the force-free equilibria --- 2D ``ABC'' field (after \citet{Arnold:1965aa}, \citet{Beltrami:1902aa}, \citet{Childress:1970aa}, e.g. \citet{Moffatt:1986aa})--- to explore the viability of
impulsive magnetic energy release. They reach the same conclusion that these configurations are unstable to ideal modes and collapse over dynamic time scales. They show that when the structures are initially compressed or sheared (driven system), the development of the instability can be accelerated. 

Although the above plasma configurations are quite artificial, they teach us a lot about the generic behavior of highly magnetized, relativistic plasmas. Similarly complex, current-carrying plasma states with plenty of free energy may form in the pulsar wind \citep[e.g.,][]{Zrake:2016aa}, or downstream of the oblique termination shock in PWN where the magnetization is still high \citep{Lyutikov:2016aa}, and also in jets due to kink instability. The insights gained here are instructive to envisage the possibility of catastrophic conversion of large scale electromagnetic energy into particle kinetic energy.

\section{Kinetic Description}\label{sec:kinetic}
\subsection{Distribution Function}
In order to understand what are the underlying dissipation mechanisms, how the released electromagnetic energy is partitioned among the particles, and what emission the accelerated particles produce, we need to follow the evolution of particle distribution in phase space. For relativistic plasma, we use the following covariant definition of distribution function over the 6-dimensional phase space: $F(\pmb{x},\pmb{u},t)=dN/(d^3x\,d^3u)$, where $\pmb{x}$ is the position and $\pmb{u}$ is the spatial part of the 4-velocity of the particles. In many of the aforementioned astrophysical environments, the plasma is collisionless, meaning that the Coulomb collisional time scale is much longer than the relevant dynamic time scales. The evolution of the distribution function for each species $F_s$ then follows the collisionless Boltzmann equation, or Vlasov equation:
\begin{equation}
\frac{\partial F_s}{\partial t}+\nabla_{\pmb{x}}\cdot(\pmb{v}F_s)+\nabla_{\pmb{u}}\cdot(\frac{d\pmb{u}}{dt}F_s)=0,
\end{equation}
where $\pmb{v}=\pmb{u}/\gamma$ is the 3-velocity, $\gamma=\sqrt{1+u^2/c^2}$ is particle Lorentz factor, and the acceleration of individual particles is determined by
\begin{equation}
m_s\frac{d\pmb{u}}{dt}=q_s(\pmb{E}+\frac{\pmb{v}}{c}\times\pmb{B}).
\end{equation}
Other forces, like radiation reaction force, can be included as well. The electromagnetic field determines the motion of particles, at the same time its evolution is determined by the charge and current density provided by the particles
\begin{gather}
\rho=\sum_s q_s\int F_s d^3u,\label{eq:rho}\\
\mathbf{J}=\sum_s q_s\int F_s\frac{\mathbf{u}}{\gamma}d^3u,\label{eq:J}
\end{gather}
This is a complex system to solve. The current state-of-the-art technique is Particle-In-Cell (PIC) simulations, which exploit a fixed spatial grid to evolve the electromagnetic field, while the particle distribution function is sampled by a large number of particles. PIC method has been successfully applied to study collisionless shocks \citep[e.g.,][]{Spitkovsky:2008aa,Sironi:2009aa,Sironi:2011aa,Sironi:2011ab}, plane current sheet reconnection \citep[e.g.][]{Kagan:2015aa}, magneto-rotational instability \citep[MRI, e.g.][]{Hoshino:2015aa,Riquelme:2012aa}, global pulsar magnetospheres \citep[e.g.,][]{Chen:2014aa,Belyaev:2015aa,Cerutti:2015aa,Philippov:2015aa}, and so on, with very fruitful outcomes.

In the following, we summarize some key results from PIC simulations of the rapid energy release processes mentioned in \S\ref{subsec:MHD}. These simple yet instructive models serve as a good testbed for understanding the details of electromagnetic dissipation, particle acceleration and high energy radiation.

\subsection{Relaxation of force-free equilibria}
\begin{figure}[htb]
  \centering
        \includegraphics[width=\textwidth]{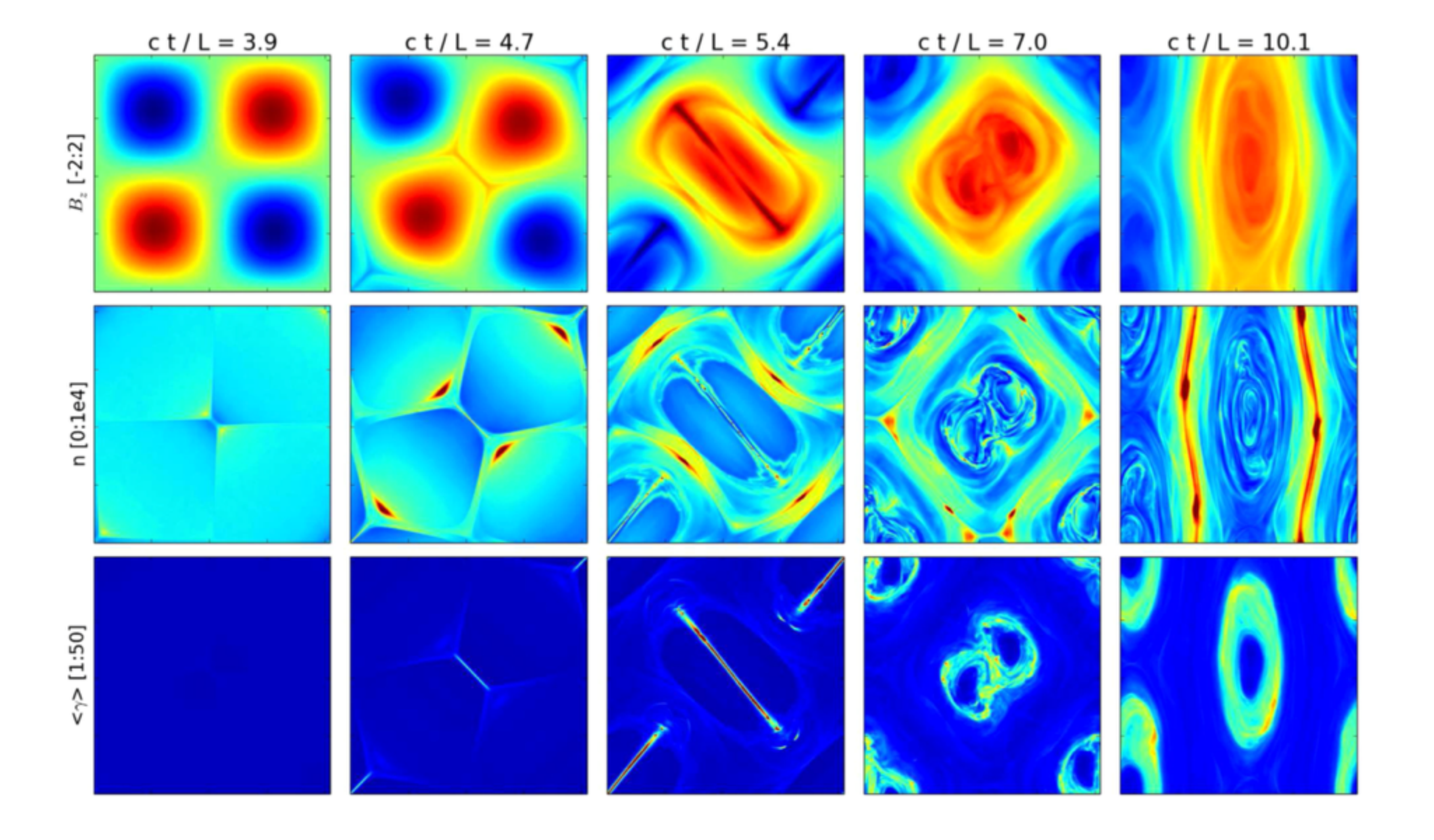}
  \caption{Snapshots from a 2D PIC simulation of one of the lowest order unstable force-free equilibria \citep[from][]{Nalewajko:2016aa}. Top panels: out of plane component of the magnetic field (there is also in-plane magnetic field); middle panels: number density $n$ of electrons and positrons; bottom panels: average Lorentz factor $\langle\gamma\rangle$ of electrons and positrons. }\label{fig:ABCevo-PIC}
\end{figure}

A few groups, including \citet{Nalewajko:2016aa,Lyutikov:2016aa,Yuan:2016aa}, have carried out 2D PIC simulations of the unstable force-free equilibria. In PIC simulations, the overall evolution of the system is quite similar to MHD/force-free simulations (Fig. \ref{fig:ABCevo-PIC}): initially the ideal instability grows on Alfv\'{e}n wave crossing time scales, producing current layers at the interface of merging flux ropes---this is consistent with the X-point collapse scenario for current sheet formation. The current layers are short-lived though, and the system enters a turbulent relaxation phase, eventually settling into the longest wavelength configuration, with almost all the available magnetic free energy converted to particle kinetic energy. The total helicity is also approximately conserved in PIC simulations.

\begin{figure}[htb]
  \centering
        \includegraphics[width=0.7\textwidth]{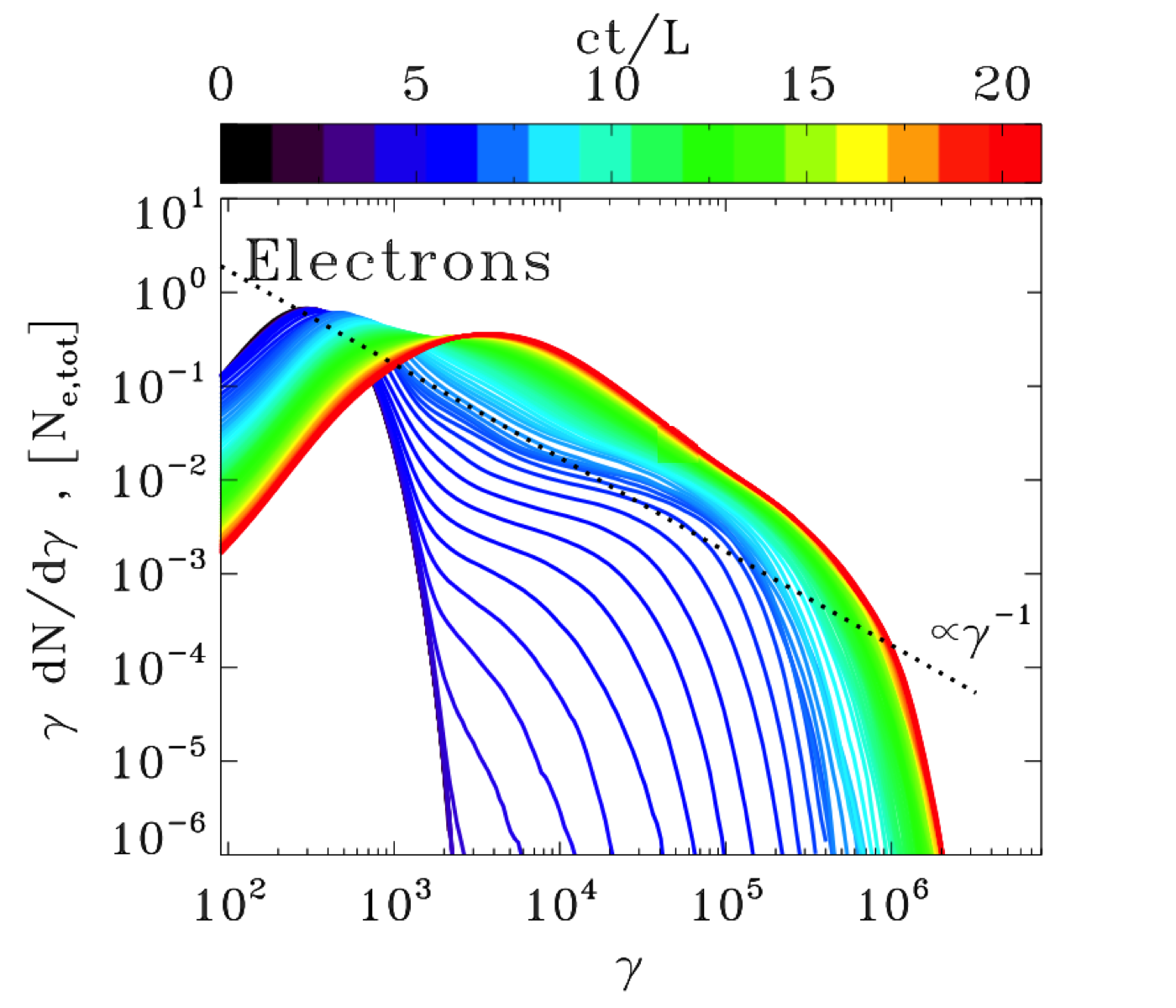}
  \caption{Particle spectrum from a 2D PIC simulation of ABC instability with initial temperature $kT/mc^2 = 10^2$, in-plane magnetization $\sigma_{\rm in} = 42$ (defined with respect to the enthalpy $w$), and $L=126r_{\rm L,\, hot}$, where L is the linear size of one flux tube, $r_{\rm L,\, hot}$ is the Larmor radius of high energy electrons heated/accelerated by reconnection \citep[from][]{Lyutikov:2016aa}. The spectrum during the initial phase of rapid acceleration (blue to cyan lines) can be harder than $\gamma dN/d\gamma \propto \gamma^{-1}$ (compare with the dotted line), as indeed required by the observations of the Crab flares.}\label{fig:ABCspectrum-PIC}
\end{figure}

However, PIC simulations show richer structures during the evolution, especially the kinetic effect on current sheet formation --- plasmoid generation during the thinning and stretching of the current sheets, and magnetic reconnection at these sheets. It is observed that the thickness of the current layer at its maximal stretch scales as the Larmor radius $r_{\rm L,\, hot}$ of the high energy electrons accelerated/heated by reconnection. The reconnection rate is measured to be in the range $v_{\rm rec}/c\sim0.2-0.5$, which increases with the magnetization and saturates at around 0.5 at high magnetization limit \citep{Lyutikov:2016aa}. There appear to be two different phases of particle acceleration in the whole process. The first phase corresponds to the formation of the first, biggest current layers where reconnection of in-plane field takes place. $\pmb{E}\cdot\pmb{B}\ne0$ in the current sheet, with the reconnection electric field $\pmb{E}$ directed primarily out of plane, which causes run-away particle acceleration in the current sheet. This acceleration is fast, producing a relatively hard, high energy component in the particle distribution (Fig. \ref{fig:ABCspectrum-PIC}). The second phase follows when the first current layers dissolve and the system evolves chaotically; the oscillating field structures scatter the particles around; there is evidence of second order Fermi acceleration smoothing out the high energy component to a softer power law, also bulk heating that increases the temperature of the thermal component.

It turns out that the particle acceleration efficiency depends on the mean magnetization of the configuration, in a similar fashion as plane current sheet reconnection scenarios \citep{Guo:2014aa,Sironi:2014aa,Werner:2016aa}. The particle spectrum gets harder as the mean magnetization increases; both the non-thermal particle fraction and the maximum particle energy increase with the magnetization \citep{Lyutikov:2016aa,Nalewajko:2016aa}.
\begin{figure}[htb]
  \centering
        \includegraphics[width=\textwidth]{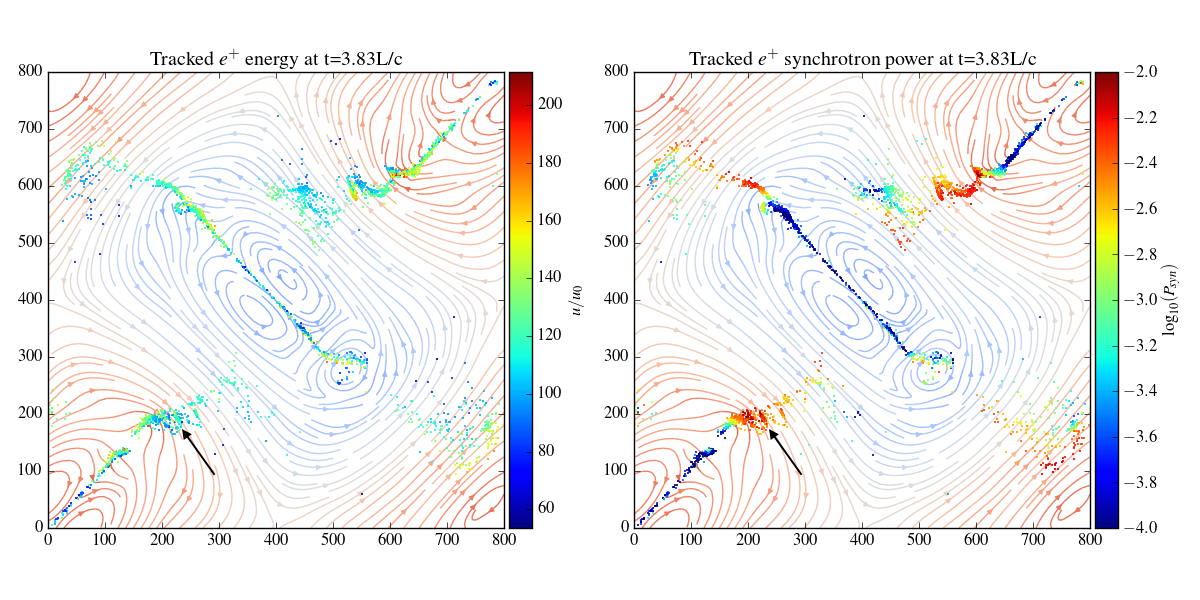}
  \caption{A snapshot of the location of tracked high energy positrons, plotted over the instantaneous field structure \citep{Yuan:2016aa}. In the left panel, the particles are color-coded by their energy while in the right panel they are color-coded by the synchrotron power. The arrows locate the ejection of plasmoids.}\label{fig:ABCrad-PIC}
\end{figure}

When we look at the synchrotron radiation signals from this evolving plasma, we find that, since the highest energy particles are first accelerated in the current layers by the parallel electric field, they do not radiate much when they are inside the sheet, because the curvature of their trajectory is small (despite the presence of guide field in the current layer). Most of the radiation is produced when particles are ejected from the current layers---their trajectories start to bend significantly in the ambient magnetic field which changes direction at the end of the current layer (Fig. \ref{fig:ABCrad-PIC}). Such a separation of acceleration site and radiative loss site could in principle facilitate acceleration beyond the synchrotron radiation reaction limit \citep{Uzdensky:2011aa}. 
Fast variability of observed photon flux can be produced when compact plasmoids that contain high energy particles are ejected from the ends of the current layers and get destroyed. These give beamed radiation. An observer sees high intensity radiation when the beam happens to be aligned with the line of sight. This is a picture similar to the kinetic beaming seen in plane current sheet reconnection simulations \citep{Cerutti:2013aa,Cerutti:2014aa}. As a result, the high energy radiation is much more variable than the low energy radiation, and these emission peaks are accompanied by an increase in the polarization degree and rapid change of polarization angle in the high energy band \citep{Yuan:2016aa}. The variability timescale is determined by the spatial extent of the emitting structure, e.g. the plasmoids, thus can be much shorter than the light crossing time of the region that collapses.

\subsection{X-point collapse}
\begin{figure}[!ht]
 \centering
\includegraphics[width=.79\textwidth]{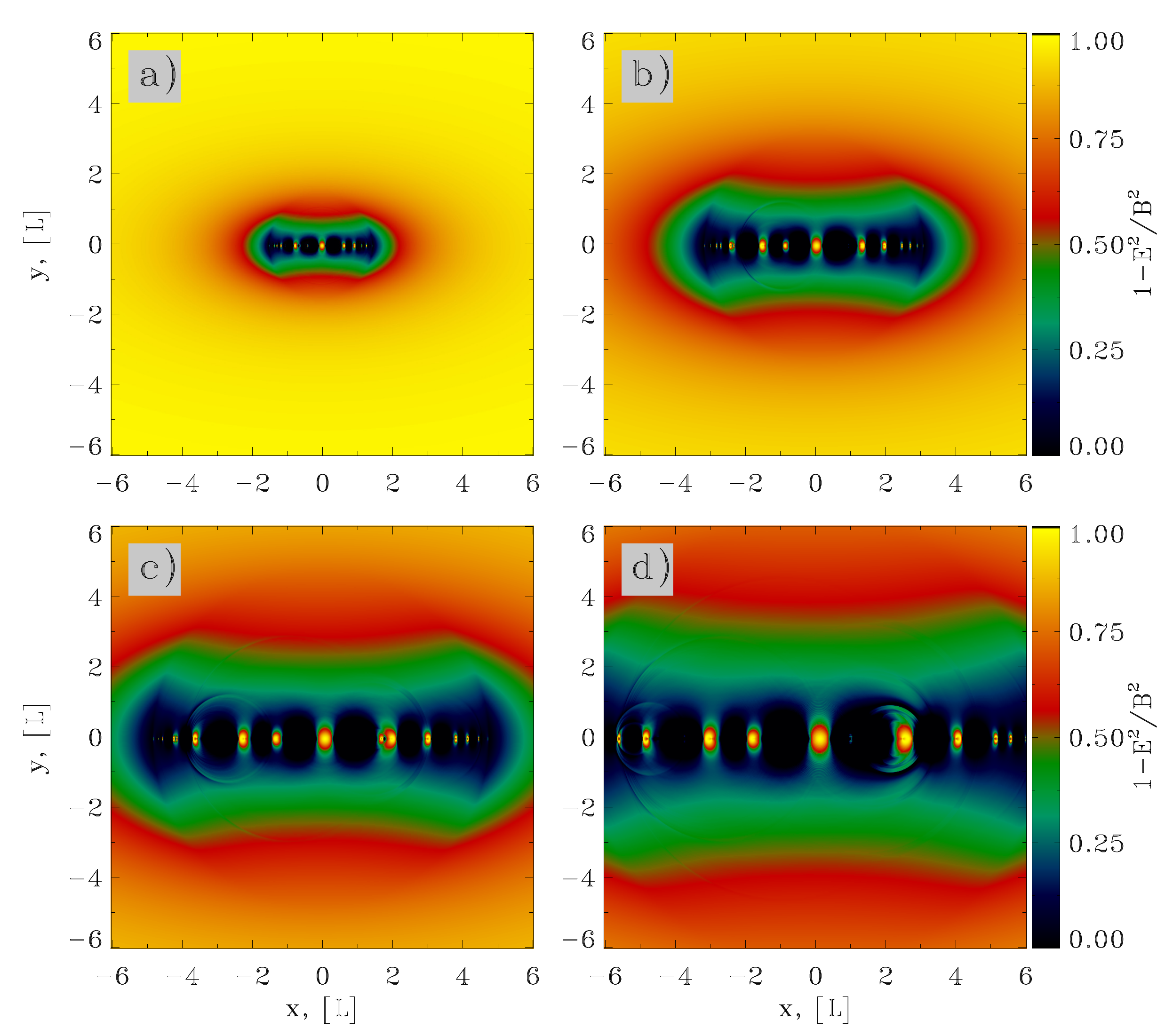} 
\caption{Late time evolution of the X-point collapse in PIC simulations with zero guide field \citep{Lyutikov:2016aa}, for $\sigma =4\times 10^{4}$  (the magnetization is measured in the initial setup at unit distance from the center). The plots show the quantity $1-E^2/B^2$ (strictly speaking, we plot $\max[0,1-E^2/B^2]$) at $ct/L=$1.5, 3, 4.5, 6, from panel (a) to (d).}
\label{fig:lor1} 
\end{figure}

As we have discussed above, the initial rapid phase of particle acceleration in 2D PIC simulations of  unstable force-free equilibria occurs at the current sheets created in between flux ropes. There, the evolution of the field line geometry resembles the X-point collapse studied by Syrovatsky in the non-relativistic regime \citep{Syrovatskii:1966aa,Imshennik:1967aa} and by \citet{Lyutikov:2016aa} in the relativistic regime. The vector potential for the X-point geometry is $A_z=-1/2\,(x^2/a(t)^2-y^2/b(t)^2)$, where $b(t)=\lambda/a(t)$. The unstressed case of $\lambda=1$ is stable. Below, we present results for the stressed case $\lambda=1/\sqrt{2}$, for magnetizations $\sigma\gg1$ and magnetic fields initialized only in the simulation plane (i.e., in the absence of the so-called ``guide field''). As shown in Fig. \ref{fig:lor1}, the collapse proceeds self-similarly:
the macroscopic distribution of $E^2/B^2$ (and hence that of the drift velocity) at later times is a scaled copy of that at previous times, with the overall length scale increasing linearly with time (at the speed of light). This implies that the reconnection rate over the whole configurations remains fixed in time (and we find that for highly magnetized plasmas, the reconnection rate $v_{\rm rec}$ approaches the speed of light on macroscopic scales).

\begin{figure}[!ht]
 \centering
\includegraphics[width=.69\textwidth]{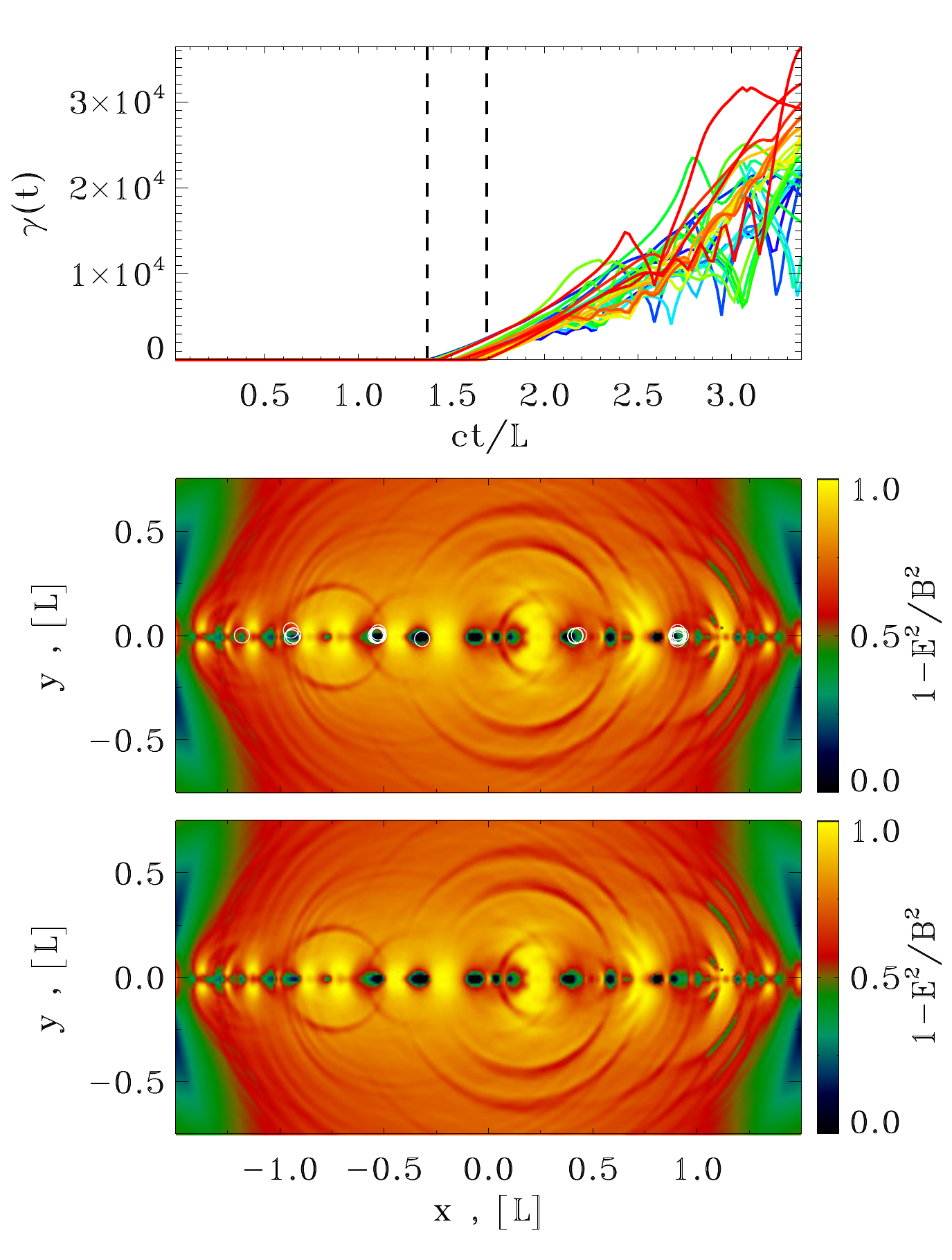} 
\caption{Physics of particle injection into the acceleration process, from a PIC simulation of stressed X-point collapse with vanishing guide field and $\sigma=4\times10^2$  \citep{Lyutikov:2016aa}. Top panel: we select all the particles that exceed the threshold $\gamma_{\rm 0}=30$ within a given time interval (chosen to be $1.4\leq ct_0/L\leq1.7$, as indicated by the vertical dashed lines), and we plot the temporal evolution of the Lorentz factor of the 30 particles that at the final time reach the highest energies. The particle Lorentz factor increases as $\gamma\propto t^2-t_0^2$, where $t_0$ marks the onset of acceleration (i.e., the time when $\gamma$ first exceeds $\gamma_{\rm 0}$). Middle panel: for the same particles as in the top panel, we plot their locations at the onset of acceleration with open white circles, superimposed over the 2D plot of $1-E^2/B^2$ (more precisely, of $\max[0,1-E^2/B^2]$). Comparison of the middle panel with the bottom panel shows that particle injection is localized in the vicinity of the X-points in the current sheet (i.e., the blue regions where $E>B$).}
\label{fig:xaccfluid} 
\end{figure}

{Interestingly, the electric field increases linearly with time. This is ultimately a manifestation of the self-similar macroscopic evolution of the system.  Indeed,  since in the initial configuration the magnetic field strength grows linearly with distance from the origin (i.e., the center of the X-point) and the current sheet size grows linearly with time, the mean magnetic and electric fields in the volume surrounding the current sheet must also grow linearly, with their scaled distributions unchanged. }
The temporal evolution of the electric field has a direct impact on the maximum particle energy. Quite generally, its time evolution will be
\be\label{eq:ggmax2}
\gamma_{\rm max} \propto E t\propto v_{\rm rec} B t
\ee
Since both $E$ and $B$ in the reconnection region are scaling linearly with time, one expects $\gamma_{\rm max}\propto t^2$. This is confirmed by \fig{xaccfluid}, where we  follow the trajectories of a number of particles in a simulation with $\sigma=4\times10^2$. The particles are selected such that their Lorentz factor exceeds a given threshold $\gamma_0=30$ within the time interval $1.4\leq ct_0/L\leq1.7$, as indicated by the vertical dashed lines in the top panel. The temporal evolution of the Lorentz factor of such particles, presented in the top panel for the 30 positrons reaching the highest energies, follows the track $\gamma\propto t^2-t_0^2$ that is expected from $d\gamma/dt\propto E(t)\propto t$. Here, $t_0$ is the injection time, when the particle Lorentz factor $\gamma$ first exceeds the threshold $\gamma_{\rm 0}$. The individual histories of single positrons might differ substantially, but overall the top panel of \fig{xaccfluid} suggests that the acceleration process is dominated by direct acceleration by the reconnection electric field.
We find that the particles presented in the top panel of \fig{xaccfluid} are too energetic to be confined within the small-scale plasmoids in the current sheet (see the small scale structures in \fig{lor1}), so any acceleration mechanism that relies on plasmoid mergers is found to be unimportant, in this setup.

Particle injection into the acceleration process happens in the charge-starved regions where $E>B$, i.e., in the small-scale X-points that separate each pair of secondary plasmoids in the current sheet. Indeed, for the same particles as in the top panel, the middle panel in \fig{xaccfluid} presents their locations at the onset of acceleration with open white circles, superimposed over the 2D plot of $1-E^2/B^2$ (more precisely, of $\max[0,1-E^2/B^2]$). Comparison of the middle panel with the bottom panel shows that particle injection is localized in the vicinity of the small-scale X-points in the current sheet (i.e., the blue regions where $E>B$). Despite occupying a relatively small fraction of the overall volume, such regions are of paramount importance for particle acceleration. 

The explosive stage of X-point collapse produces non-thermal tails (in analogy to the relaxation of unstable force-free structures) whose hardness depends on the average magnetization. For sufficiently high magnetizations and vanishing guide field, the non-thermal particle spectrum consists of two components: a low-energy population with soft spectrum, that dominates the number census; and a high-energy population with hard spectrum, that possesses all the properties needed to explain the Crab flares  \citep{Lyutikov:2016aa}. The particle distribution is highly anisotropic, with high-energy particles  beamed primarily along the direction of the accelerating electric field.

\subsection{Merging Lundquist flux ropes}
Above, we have described the evolution of unstable force-free configurations (with the ABC geometry being a special case). There are two key features of the preceding model that are specific to the initial set-up: (i) each flux tube carries non-zero poloidal current; (ii) the initial configuration is an unstable equilibrium. However, as we now show, the evolution is generic, regardless of these conditions. \citet{Lyutikov:2016aa} investigated a merger of two flux tubes with zero total current, with MHD and PIC simulations. Thus, two flux tubes are not attracted to each other --- at least initially. Here, we present the results for Lundquist's force-free cylinders surrounded by uniform magnetic field,
\be
{\bf B}_L(r\le r_{\rm j}) = J_1 (r \alpha) {\bf e}_\phi +  J_0 (r \alpha) {\bf e}_z\,,
\label{eq:lundquist}
\ee
Here, $J_{0},\,J_{1}$ are Bessel functions of zeroth and first order and the constant $\alpha\simeq 3.8317$ is the first root of $J_{0}$.  
We chose to terminate this solution at the first zero of $J_1$, which we denote 
as $r_j$ and hence continue with $B_z=B_z(r_j)$ and $B_\phi=0$ for $r>r_j$. Thus the total current of the flux tube is zero.
As the result, the azimuthal field vanishes at the boundary of the rope, whereas the poloidal one changes sign inside the rope. The evolution is very slow, given the fact that at the contact the reconnecting field vanishes (i.e., the initial configuration is dynamically stable).
 To speed things up, the ropes are pushed towards each other.
 
 \begin{figure}[!]
\centering
\includegraphics[width=.34\textwidth]{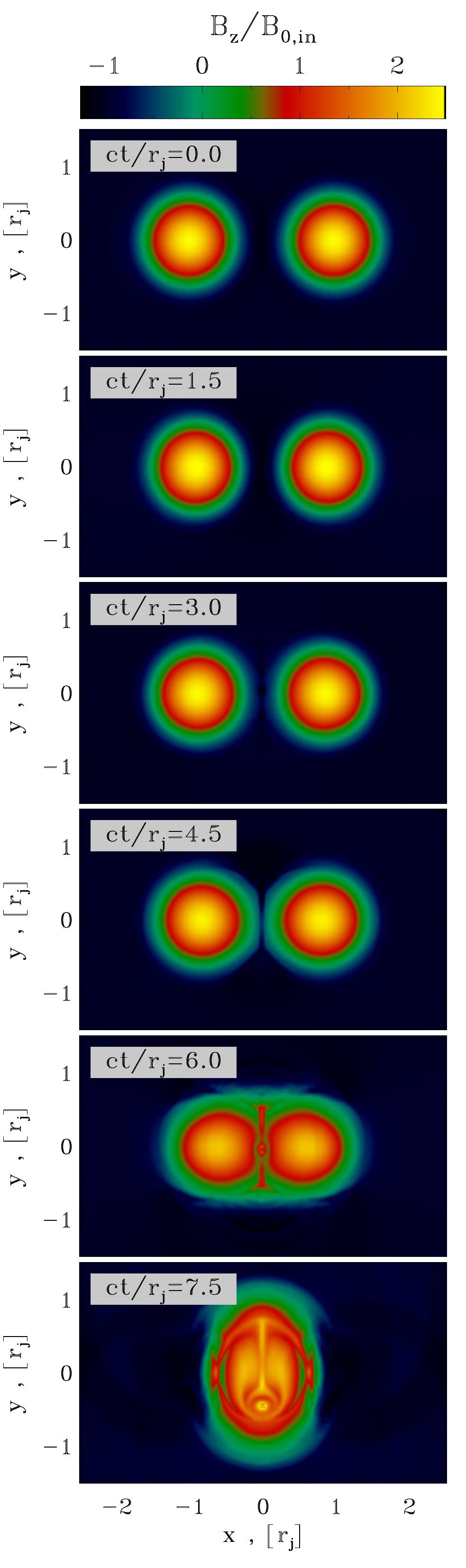} 
\includegraphics[width=.34\textwidth]{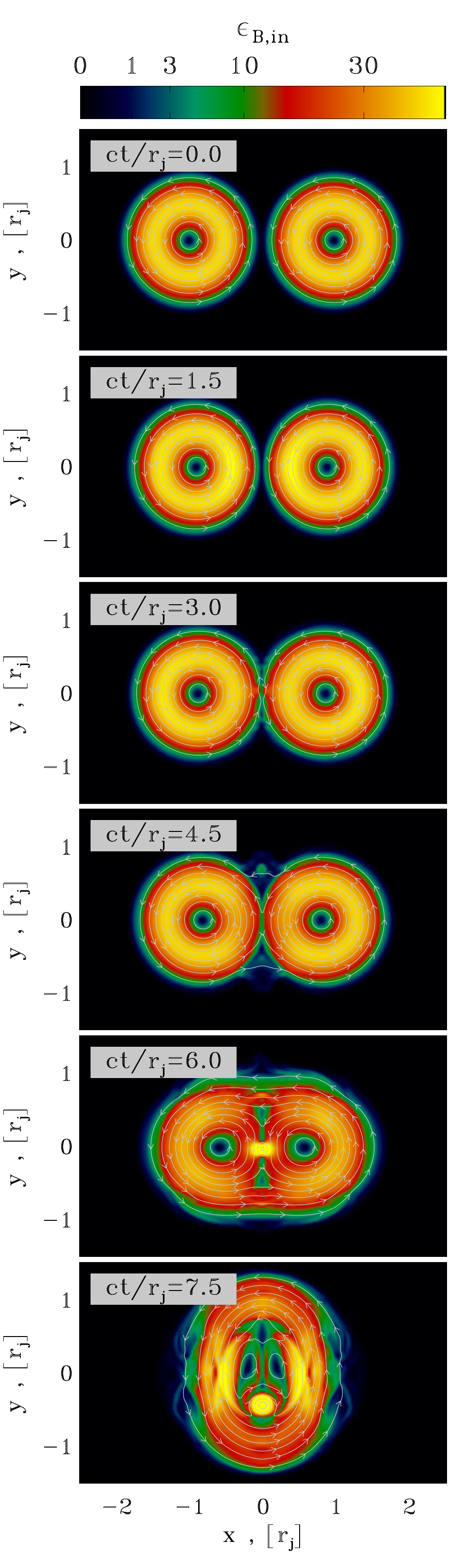} 
\caption{Temporal evolution of 2D Lundquist ropes (time is measured in $c/\rj$ and indicated in the grey box of each panel, increasing from top to bottom), from  \citet{Lyutikov:2016aa}. The plot presents the 2D pattern of the out-of-plane field $B_z$ (left column) and of the in-plane magnetic energy fraction $\epsilon_{B,\rm in}=(B_x^2+B_y^2)/8 \pi n m c^2$ (right column; with superimposed magnetic field lines), from a PIC simulation with $kT/mc^2=10^{-4}$, $\sigmain=43$ and $\rj=61\,\rhot$.}
\label{fig:lundfluid} 
\end{figure}
\begin{figure}[h!]
\centering
\includegraphics[width=.54\textwidth]{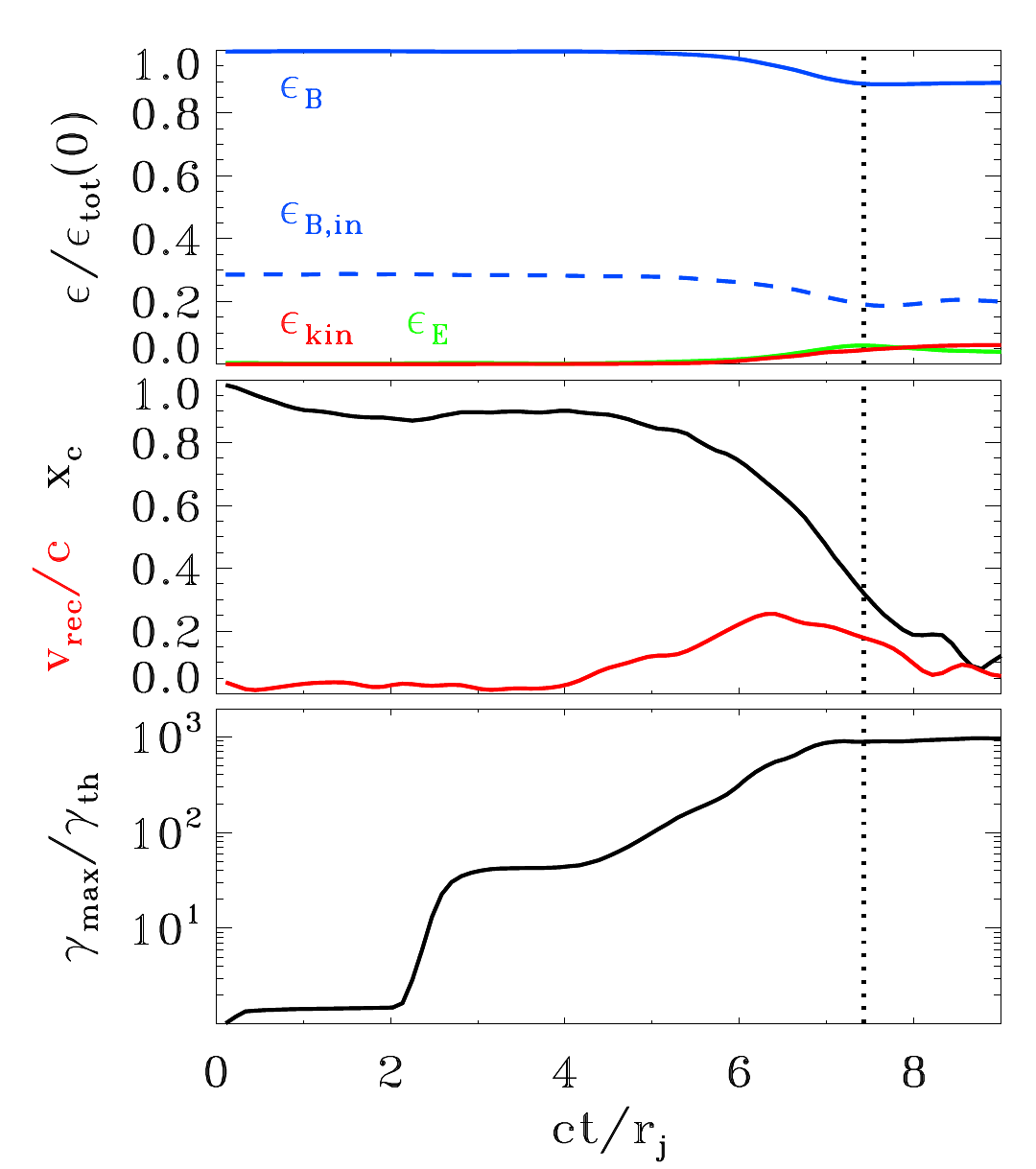} 
\caption{Temporal evolution of various quantities, from a 2D PIC simulation of Lundquist ropes with $kT/mc^2=10^{-4}$, $\sigmain=43$ and $\rj=61\,\rhot$ (the same as in \fig{lundfluid}), from  \citet{Lyutikov:2016aa}. Top panel: fraction of energy in magnetic fields (solid blue), in-plane magnetic fields (dashed blue), electric fields (green) and particles (red; excluding the rest mass energy), in units of the total initial energy. Middle panel: reconnection rate $v_{\rm rec}/c$ (red), and location $x_{\rm c}$ of the core of the rightmost flux rope (black), in units of $\rj$.
Bottom panel: evolution of the maximum Lorentz factor $\gammamax$.}
\label{fig:lundtime} 
\end{figure}
 
In \fig{lundfluid}, we present the 2D pattern of the out-of-plane field $B_z$ (left column) and of the in-plane magnetic energy fraction $\epsilon_{B,\rm in}=(B_x^2+B_y^2)/8 \pi n m c^2$ (right column; with superimposed magnetic field lines), from a PIC simulation with $kT/mc^2=10^{-4}$, $\sigmain=43$ (only defined with the in-plane fields) and $\rj=61\,\rhot$.
As the two magnetic ropes slowly approach, driven by the initial velocity push, reconnection is triggered in the plane $x=0$, as indicated by the formation and subsequent ejection of small-scale plasmoids. Until $ct/\rj\sim 4.5$, the cores of the two islands have not significantly moved (black line in the middle panel of \fig{lundtime}, indicating the $x_{\rm c}$ location of the center of the rightmost island), the reconnection speed is quite small (red line in the middle panel of \fig{lundtime}) and no significant energy exchange has occurred from the fields to the particles (compare the in-plane magnetic energy, shown by  the dashed blue line in the top panel of \fig{lundtime}, with the particle kinetic energy, indicated with the red line). 

As a result of reconnection, an increasing number of field lines, that initially closed around one of the ropes, are now engulfing both magnetic islands. Their tension force causes the two ropes to approach and merge on a quick (dynamical) timescale, starting at $ct/\rj\sim 4.5$ and ending at $ct/\rj\sim 7.5$ (see that the distance of the rightmost island from the center rapidly decreases, as indicated by the black line in the middle panel of \fig{lundtime}). The tension force drives the particles in the flux ropes toward the center, with a fast reconnection speed peaking at $v_{\rm rec}/c\sim 0.3$ (red line in the middle panel of \fig{lundtime}). The reconnection layer at $x=0$  stretches up to a length of $\sim 2\rj$, and secondary plasmoids are formed. In the central current sheet, it is primarily the in-plane field that gets dissipated (compare the dashed and solid blue lines in the top panel of \fig{lundtime}), driving an increase in the electric energy (green) and in the particle kinetic energy (red). In this phase of evolution, the fraction of initial energy released to the particles is small ($\epsilon_{\rm kin}/\epsilon_{\rm tot}(0)\sim 0.1$), but the particles advected into the central X-point experience a dramatic episode of acceleration. As shown in the bottom panel of \fig{lundtime}, the cutoff Lorentz factor $\gammamax$ of the particle spectrum presents a dramatic evolution, increasing up to $\gammamax/\gamma_{\rm th}\sim 10^3$ within a couple of dynamical times. This phase of extremely fast particle acceleration  on a dynamical timescale is analogous to the relaxation of unstable force-free structures discussed above.

\section{Future Directions}\label{sec:future}
\subsection{Observation}
Much has been learned in recent years from observations throughout the electromagnetic spectrum (radio through TeV $\gamma$-rays) and beyond (gravitational radiation, neutrinos and cosmic rays). Even more discoveries are anticipated from observations with new facilities that should come on line in the next $\sim5$ years. Perhaps the most exciting opportunity is to make a (temporal) match of a short GRB with a LIGO event. Equally interesting would be a temporal association of an``cosmogenic'' VHE neutrino with some other cosmic event. We may also learn about the details of jet production in AGN from Event Horizon Telescope observations of M87, as discussed above. Fast Radio Bursts are starting to reveal their secrets and they, too, may be shown to have some physical affinity with the other examples of release of energy, in particular magnetars. Comparisons of radio pulsar observations with corresponding $\gamma$-ray observations and studies of radio scintillation are providing a much better understanding of neutron star magnetospheres which may help us calibrate our models of other magnetically dominant sources.  Similar remarks apply to solar flares where there can be a rapid release of magnetic energy in a very short space of time. 

There will also be new optical/near infrared survey telescopes devoted to transient astronomy including the Zwicky Transient Factory, LSST and Euclid,  greatly increasing our ability to monitor familiar types of transient, as well as discover new classes of which we are now ignorant.
\subsection{Simulation}
With the advance in algorithms, simulation techniques and computational power, much more can be understood about plasma physics under extreme conditions using simulations. On one hand, fluid level simulations are now able to handle complex, realistic astrophysical situations, like the merging of neutron stars/black holes, tidal disruption events, colliding winds in binary systems, accretion disk and jet formation, etc. These provide more physical interpretation of the observed phenomena, also point to possible sites and mechanisms of dissipation and particle acceleration. 

On the other hand, kinetic simulations are getting close to modeling multi-scale systems. For example, the role of reconnection and particle acceleration in accretion disk has been recently discussed in PIC simulations \citep[e.g.][]{Hoshino:2015aa,Riquelme:2012aa}. PIC simulations are also used to study global pulsar magnetospheres  \citep[e.g.,][]{Chen:2014aa,Belyaev:2015aa,Cerutti:2015aa,Philippov:2015aa}, relativistic turbulence \citep{Zhdankin:2017aa}, etc. Improved hybrid methods are also bridging between the kinetic and macroscopic length scales \citep{Kunz:2016aa}. A major challenge is to increase the dynamic range of particle energy that can be covered in these simulations to approach that found in the sources discussed above. Furthermore, new physics is being added to traditional PIC simulations, including pair production \citep{Chen:2014aa}, general relativistic effect \citep{Philippov:2015aa}, radiative feedback \citep{Uzdensky:2016aa}, and hadronic interactions. 

\subsection{Experiment}
Another new development is the prospect of performing novel experiments at the many powerful light sources around the world that are already operational or are expected to become such over the next five years. There are various schemes being explored. For example, $\gamma$-rays can be created by Bethe-Heitler bremsstrahlung when a relativistic electron beam interacts with a foil target. Electron-positron pairs will also be produced. However if the $\gamma$-rays are above threshold they can also pair-produce on an oncoming, coherent X-ray beam from a powerful laser by the Breit-Wheeler process, just like in cosmic sources. Achieving this pair production will be a major experimental milestone but will have no significance for QED as the cross-section is not in doubt! However, it will open the door to explorations of nonlinear quantum electrodynamics where calculations are hard and deserve validation. In addition and of even more interest for astrophysics are the many body processes where it is the collective properties of the pair plasma that are important. There are additional processes that can be investigated in the same spirit involving large magnetostatic fields. 

These explorations, although partly inspired by the challenges of extreme astrophysical acceleration, will be prosecuted in the spirit of comparing experiments that can be performed and diagnosed with numerical simulations. It is an exciting prospect that new effects will be discovered and that some of these may have a role in understanding the processes we have discussed in this chapter.      
\begin{acknowledgements}
We thank Andrei Bykov for his patience and for organizing a very interesting meeting, the other participants for helping us develop our ideas, and many other colleagues, including William East, Serguei Komissarov, Maxim Lyutikov, Krzysztof Nalewajko, Oliver Porth and Jonathan Zrake for their collaboration on research described in this chapter.
\end{acknowledgements}
\bibliographystyle{spbasic1}
\bibliography{magnetoluminescence}
\end{document}